\documentclass[12pt,preprint]{aastex}

\def\curf{{\cal F}}

\def\MSUNYR{\rm M_{\odot}\,yr^{-1}}
\def\MSUN{\rm M_{\odot}}
\def\LSUN{\rm L_{\odot}}
\def\RSUN{\rm R_{\odot}}
\def\Mdot{ \dot{M}}

\begin{document}

\shortauthors{Luhman et al.}
\shorttitle{Brown Dwarf Disk}

\title{{\it Spitzer} Identification of the Least Massive Known Brown Dwarf with 
a Circumstellar Disk}

\author{
K. L. Luhman\altaffilmark{1},
Paola D'Alessio\altaffilmark{2}, 
Nuria Calvet\altaffilmark{1}, 
Lori E. Allen\altaffilmark{1}, 
Lee Hartmann\altaffilmark{1}, 
S. T. Megeath\altaffilmark{1}, 
P. C. Myers\altaffilmark{1}, 
\& G. G. Fazio\altaffilmark{1}}

\altaffiltext{1}{Harvard-Smithsonian Center for Astrophysics, 60 Garden St.,
Cambridge, MA 02138; kluhman, ncalvet, leallen, lhartmann, tmegeath, pmyers, 
gfazio@cfa.harvard.edu.}

\altaffiltext{2}{Centro de Radiostaronom\'\i a y de Astrof\'\i sica, UNAM,
Apartado Postal 72-3 (Xangari), 58089 Morelia, Michoac\'an, M\'exico;
p.dalessio@astrosmo.unam.mx}

\begin{abstract}

Using the Infrared Array Camera (IRAC) aboard the {\it Spitzer Space Telescope},
we have obtained mid-infrared photometry of the least massive known brown
dwarf in the Chamaeleon~I star-forming region.
For this young brown dwarf, OTS~44, we have constructed a spectral energy
distribution (SED) from 0.8-8~\micron\ by combining 
the measurements at 3.6, 4.5, 5.8, and 8.0~\micron\ from IRAC with
ground-based photometry at $I$, $J$, $H$, and $K$. The resulting SED for OTS~44 
exhibits significant excess emission longward of 3~\micron\ relative 
to the SED expected from the photosphere of the brown dwarf.
We have successfully modeled the source of this excess emission in terms 
of an irradiated viscous accretion disk with $\Mdot\lesssim10^{-10}$~$\MSUNYR$. 
With a spectral type of M9.5 and a mass of $\sim15$~$M_{\rm Jup}$, 
OTS~44 is now the coolest and least massive brown dwarf observed to have 
a circumstellar disk.
These measurements demonstrate that disks exist around brown dwarfs even down
to the deuterium burning mass limit and the approximate upper mass limit of
extrasolar planetary companions.

\end{abstract}


\section{Introduction}
\label{sec:intro}

Measuring the lowest mass at which brown dwarfs harbor circumstellar disks is
crucial for understanding the formation mechanism of brown dwarfs and 
for determining whether planets can form around low-mass brown dwarfs. 
The natural laboratories for performing this measurement are the 
nearest regions of star formation ($\tau\sim0.5$-3~Myr, $d=150$-300~pc), 
particularly those that have been surveyed extensively for brown dwarfs
\citep[e.g.,][]{luh99,com00,mar01,bri02}.
For some of the young brown dwarfs discovered in these
surveys\footnote{The hydrogen burning mass limit
at ages of 0.5-3~Myr corresponds to a spectral type of $\sim$M6.25
according to the models of \citet{bar98} and \citet{cha00} and the
temperature scale of \citet{luh03b}.},
indirect evidence of circumstellar disks has been obtained through
detections of accretion. 
The most readily observed signature of accretion in a young brown dwarf
is a broad H$\alpha$ emission line profile, which has been observed through 
high-resolution spectroscopy in GM Tau, IC~348-382, and IC~348-415 at a 
spectral type of M6.5 
\citep{wb03,muz03b} and in IC348-355 and 2MASS~1207-3932 at M8 
\citep{jay03b,moh03}.
Very large equivalent widths of H$\alpha$ in low-resolution spectra also
have been suggestive of accretion in KPNO-Tau~12 and S~Ori~55 at M9 
\citep{luh03a,zap02} and in S~Ori~71 at L0 \citep{bar02}. 
High-resolution data has confirmed the presence of accretion in KPNO-Tau~12,
as well as in several other young sources at M6.25-M8.5 \citep{muz05,moh05}.

Direct evidence of circumstellar disks around brown dwarfs has been provided
by detections of infrared (IR) emission above that expected from stellar
photospheres alone. Modest excess emission in the $K$ and $L$ bands has been 
observed for several young objects at M6-M8 and for a few as late as M8.5
\citep{luh99,luh04a,lada00,lada04,mue01,liu03,jay03a}.
At longer, mid-IR wavelengths, the {\it Infrared Space Observatory (ISO)} 
detected excesses for CFHT~4 at M7 \citep{pas03}, Cha~H$\alpha$~1 at M7.75 
\citep{per00,com00,nat01b}, and GY141 at M8.5 \citep{com98}.
Photometry at similar wavelengths was recently obtained from the ground with
Gemini Observatory for CFHT~4 \citep{apa04}, Cha~H$\alpha$~1, and 
2MASS~1207-3932 \citep{ster04}. 
Detections of circumstellar material around young brown dwarfs
have been extended to millimeter wavelengths for CFHT~4, as well as 
for IC~348-613 at M8.25 \citep{kle03}.
Additional detections of accretion and disks have been reported for 
objects in Ophiuchus with IR spectral types of M6 and later 
\citep{tes02,nat02,nat04,moh04}. However, because those types
were derived from water absorption bands by using field M dwarfs as the
standards, they likely have systematic errors such that the temperatures,
and thus the masses, are underestimated \citep{luh03b,luh04b}.

To continue to search for circumstellar disks around brown dwarfs at later 
types and lower masses, we have obtained mid-IR photometry with 
the Infrared Array Camera \citep[IRAC;][]{faz04} on the 
{\it Spitzer Space Telescope} for source 44 from 
\citet{ots99} (hereafter OTS~44), which is the coolest and least massive known 
brown dwarf in the Chamaeleon~I star-forming region 
\citep[M9.5, $M\sim15$~$M_{\rm Jup}$;][]{luh04b}.
In this letter, we describe these observations, construct a spectral
energy distribution (SED) for OTS~44,
measure its mid-IR excess emission, and compare this excess to our model 
predictions for emission from a circumstellar disk.

\section{Observations}
\label{sec:obs}

As a part of the Guaranteed Time Observations of the IRAC instrument team,
we obtained images at 3.6, 4.5, 5.8, and 8.0~\micron\ with IRAC on the 
{\it Spitzer Space Telescope} of the northern cluster in the Chamaeleon~I
star-forming region on 2004 July 4 (UT).
The plate scale and field of view of IRAC are $1\farcs2$ and 
$5\farcm2\times5\farcm2$, respectively. The camera produces images with 
FWHM$=1\farcs6$-$1\farcs9$ from 3.6 to 8.0~\micron\ \citep{faz04}.
The cluster was mapped with a $6\times7$ mosaic of pointings 
separated by $280\arcsec$ and aligned with the array axes.  
At each cell in the map, images were obtained in the 12~s high dynamic 
range mode, which provided one 0.4~s exposure and one 10.4~s exposure. 
The map was performed twice with offsets of several arcseconds between 
the two iterations. The resulting maps had centers of 
$\alpha=11^{\rm h}09^{\rm m}26^{\rm s}$, $\delta=-76\arcdeg36\arcmin26\arcsec$ 
(J2000) for 3.6 and 5.8~\micron\ and at
$\alpha=11^{\rm h}10^{\rm m}16^{\rm s}$, $\delta=-76\arcdeg30\arcmin20\arcsec$ 
(J2000)
for 4.5 and 8.0~\micron, dimensions of $33\arcmin\times29\arcmin$, and 
position angles of $28\arcdeg$ for the long axes. 
The images from the Spitzer Science Center pipeline (version S10.5.0) 
were combined into one
mosaic at each of the four bands using custom IDL software developed by 
Robert Gutermuth. 

Aperture photometry for OTS~44 was extracted with the task PHOT under the 
IRAF package APPHOT using a radius of two pixels ($2\farcs4$) for the aperture 
and inner and outer radii of two and six pixels for the sky annulus. 
In the next section, we will compare the IRAC measurements of OTS~44 to those
of KPNO-Tau~4, a young brown dwarf in Taurus with the same spectral type as 
OTS~44. Therefore, we also measured KPNO-Tau~4 from images 
obtained by \citet{har05} with the same aperture sizes applied to OTS~44. 
Because the 8.0~\micron\ images of KPNO-Tau~4 were contaminated by residual
images from a prior observation of a bright star, the photometry in this band
is uncertain and therefore is not considered in this work.
For an aperture radius of 10 pixels and a sky annulus extending from 10 to 20
pixels, we 
adopted zero point magnitudes ($ZP$) of 19.601, 18.942, 16.882, 17.395 in
the 3.6, 4.5, 5.8 and 8 $\mu$m bands, where $ M = -2.5 \log (DN/sec) + ZP$.
We then applied aperture corrections of 0.210, 0.228, 0.349, and 0.499~mag 
to the photometry of OTS~44 and KPNO-Tau~4.
The resulting IRAC photometry for OTS~44 and KPNO-Tau~4 are listed in 
Table~\ref{tab:data}.
The photometric uncertainties for both objects are 0.02~mag at
3.6 and 4.5~\micron\ and 0.04~mag at 5.8 and 8.0~\micron, with an additional
10\% uncertainty in the IRAC calibration.
Table~\ref{tab:data} includes optical and near-IR photometry for OTS~44
and KPNO-Tau~4. The data at $I$ are from K. Luhman (in preparation) and 
\citet{bri02} and the near-IR measurements are from \citet{ots99} and
the 2MASS Point Source Catalog for OTS~44 and KPNO-Tau~4, respectively.

\section{Analysis}

\begin{figure}
\epsscale{0.7}
\plotone{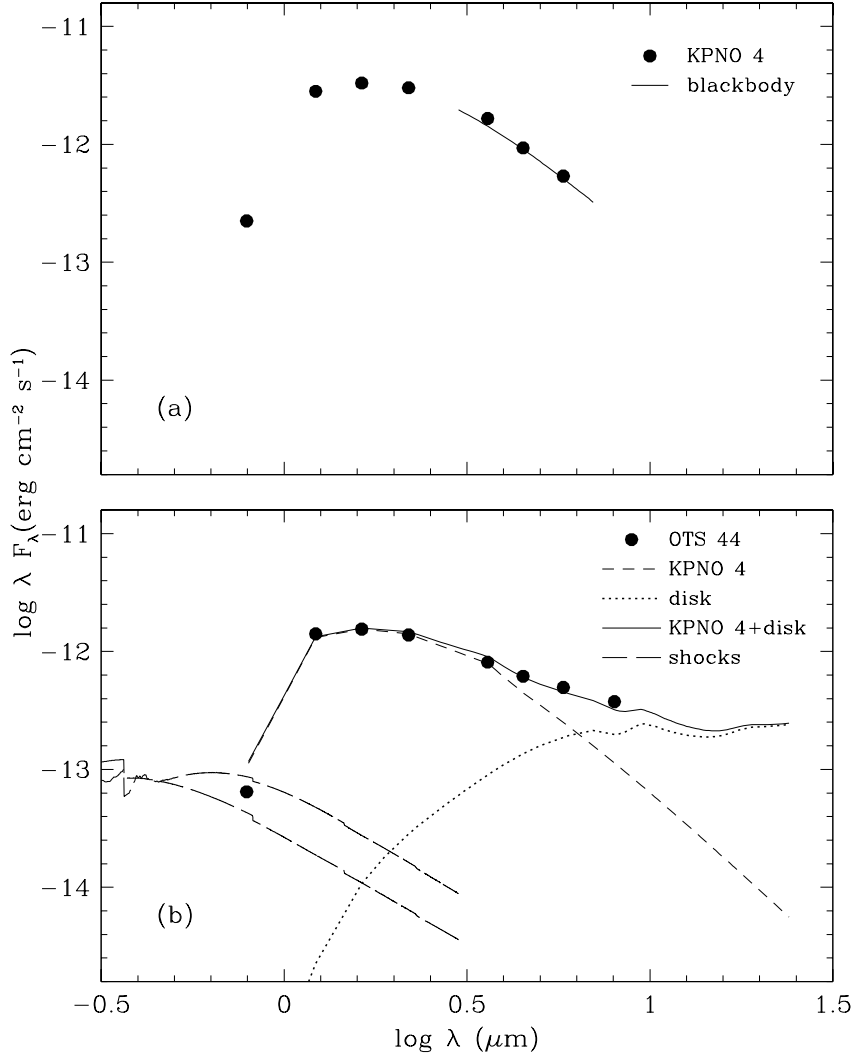}
\caption{ 
SEDs of the young brown dwarfs KPNO-Tau~4 and OTS~44 ({\it points}).
(a) The slope of the mid-IR SED of KPNO-Tau~4 is reproduced by a blackbody at 
the effective temperature of its stellar photosphere 
($T_{\rm eff}=2300$~K, {\it solid line}), 
and thus exhibits no excess emission from circumstellar material.
(b) To estimate the SED of the stellar photosphere of OTS~44, 
the measured SED of KPNO-Tau~4 is combined with the blackbody SED at 
$\lambda>6$~\micron\ and scaled to the $H$-band flux of OTS~44
({\it short dashed line}).
The excess flux above this photospheric SED is modeled in terms
of emission from a circumstellar accretion disk ({\it dotted line}). 
The sum of this model disk SED and the photospheric SED agrees fairly well
with the data ({\it solid line}).
The values of the model parameters are $\Mdot=10^{-10} \ \MSUNYR$,
an inclination angle to the line of sight $i=0^\circ$ (i.e., pole-on),
$\alpha=0.01$, and $R_{wall}=3 \ R_*$.
We also include the SEDs predicted for the accretion shock for values of the 
energy flux that are representative of higher mass T Tauri stars, 
$\curf = 10^{11}$ and $10^{12}\, {\rm erg \, cm^{-2} \, s^{-1}}$ 
({\it long dashed lines}).
}
\label{fig:sed}
\end{figure}

Before discussing OTS~44, we first examine the SED of the comparison source,
KPNO-Tau~4.
Figure~\ref{fig:sed} shows the SED of KPNO-Tau~4 derived from the
photometry in Table~\ref{tab:data} using a distance modulus of 5.76 for Taurus
\citep{wic98} and the appropriate zero-magnitude fluxes 
\citep{bes79,bb88,faz04}.
No correction for extinction was applied to these data 
because of the negligible extinction toward this object \citep{bri02}.
As shown in Figure~\ref{fig:sed}, the mid-IR SED of KPNO-Tau~4 is
consistent with a blackbody at the effective temperature of the central
object \citep[$T_{\rm eff}=2300$~K;][]{luh04b}\footnote{The mid-IR SED of 
an object at the age and temperature of KPNO-Tau~4 is predicted to agree
with that of a blackbody \citep{all00}.}, and thus exhibits no evidence of
excess emission from circumstellar material. Therefore, in the following 
analysis, we adopt the SED of KPNO-Tau~4 to represent the intrinsic photosphere 
of young objects at the spectral type of OTS~44.

We now examine the SED of OTS~44, which is shown in Figure~\ref{fig:sed}
after an extinction correction of $A_J=0.3$ \citep{luh04b} and adopting
a distance modulus of 6.13 for Chamaeleon~I \citep{whi97,wic98,ber99}.
In comparison to KPNO-Tau~4, OTS~44 exhibits excess flux in the IRAC bands
that becomes stronger with longer wavelengths. 
We have compared this excess emission with the predictions of a model of
an irradiated accretion disk. For these calculations, we adopted
a stellar mass $M_* = 0.015$~$\MSUN$ and radius $R_* = 0.23$~$\RSUN$ 
\citep{luh04b}, uniform grain size distribution everywhere given by 
the standard power law  $n(a) \sim a^{-3.5}$ with minimum and maximum grain 
sizes $a_{min}=0.005 \ \mu$m and $a_{max}=0.25 \ \mu$m \citep{mat77}, 
uniform mass accretion rate, and an inner radius $R_{in}=3$~$R_*$ 
(given by the silicate sublimation radius).
The disk was heated by local viscous dissipation and stellar irradiation. 
The emission from a wall at the dust destruction radius was included.
Detailed descriptions of the method used to calculate the
disk structure and emergent intensity are given by \citet{dal98},
\citet{dal99}, and \citet{dal01}.

As shown in Figure~\ref{fig:sed}, 
we were able to fit the observed SED with a disk model 
seen pole-on ($i=0$) with an accretion rate of $\Mdot \sim 10^{-10} \MSUNYR$ 
and an inner radius of 3~$R_*$.  We were driven to low inclinations because we 
found that any significant contribution from the inner wall, which radiates 
approximately as a blackbody at 1400 K, resulted in an SED that was 
inconsistent with the IRAC observations.
Our estimate of the accretion rate is higher than values derived for
other young brown dwarfs from modeling of H$\alpha$ 
emission profiles \citep{muz03b,muz05}.
Within the context of our model and 
the adopted inclination, we required substantial viscous dissipation due to 
accretion as well as irradiation from the central star to achieve high enough 
temperatures at the inner edge of the disk.  Our model assumes that the inner 
disk wall is perfectly vertical (perpendicular to the disk plane).  
If the disk wall is actually somewhat tilted or curved, as was adopted 
to model the older T Tauri star TW Hya \citep{cal02}, also seen pole-on, it is 
possible that the intermediate range of temperatures at the inner disk
edge resulting from such structure would enable us to fit the observations with
lower $\Mdot$.  Further exploration of this issue is needed.

The accretion luminosity for the adopted values of the stellar mass 
and radius and the disk mass accretion rate is 0.0002~$\LSUN$, which 
corresponds to 15\% of the stellar luminosity, 
$L_* = 0.0013$~$\LSUN$ \citep{luh04b}.
If matter from the disk is loaded onto the
star through a magnetospheric flow, as the observed
line profiles in higher mass brown dwarfs and very low mass stars seem to 
indicate \citep{muz03b,muz05,wb03,jay03b,moh03,moh05},
then an accretion shock should form at the stellar
surface, where most of the accretion luminosity
would be emitted \citep{cg98,gul00}.
This emission would veil the photospheric lines.
To estimate the expected degree of veiling, 
we have calculated the emission from accretion
shocks on the stellar surface following the procedures
of \citet{cg98}. Figure~\ref{fig:sed} shows the shock emission
from two models of accretion columns with values of
the energy flux in the column 
$\curf = 10^{11}\, {\rm erg \, cm^{-2} \, s^{-1}}$
and $\curf = 10^{12}\, {\rm erg \, cm^{-2} \, s^{-1}}$, which are 
representative of those estimated for higher mass T Tauri stars
\citep{cg98} and correspond to filling factors
on the stellar surface of 0.0016 and 0.00016, respectively.
As shown in Figure~\ref{fig:sed}, veiling from the shock is predicted 
to be significant at optical wavelengths but negligible longward of 1~\micron. 
Optical spectroscopy can test the model prediction of a high accretion rate;
as discussed above, modifying the structure of the inner disk edge could result
in acceptable fits with lower accretion rates and thus lower accretion shock
luminosities.

The IRAC data do not constrain the maximum grain size.
Disk models with different grain maximum sizes have the same SED
for $\lambda < 9 \ \mu$m because the fluxes arise from the optically
thick and flat inner disk regions.
Thus, observations at longer wavelengths are necessary 
for constraining the dust properties in the OTS 44 disk. Similarly, it is not 
possible to constrain the disk mass with the available data; 
models with a different viscosity parameter $\alpha$ and disk radius 
(which implies different disk masses for a given mass accretion rate)
have the same SED in the observed spectral range. 

\section{Discussion}

Through observations with IRAC aboard the {\it Spitzer Space Telescope},
we have shown that the least massive known member of Chamaeleon~I, OTS~44, 
exhibits strong mid-IR excess emission, which we have reproduced with
a model of an accretion disk. With a spectral type of M9.5 and a mass of 
$\sim15$~$M_{\rm Jup}$, 
OTS~44 is now the coolest and least massive brown dwarf observed to have 
a circumstellar disk. The presence of a disk around OTS~44 demonstrates
that the formation of free-floating bodies via disks extends down to the
deuterium burning mass limit \citep[$M\approx14$~$M_{\rm Jup}$;][]{bur97}, 
and raises the possibility of planet formation around objects that themselves
have planetary masses. 
Indeed, a candidate planetary companion has already been discovered 
near the earlier and more massive young brown dwarf 2MASS~1207-3932 
\citep[M8, $M\sim30$~$M_{\rm Jup}$;][]{giz02} by \citet{chau04}.

The relative ease of the detection of the excess emission from OTS~44
($\tau_{\rm int}=20.8$~s) 
demonstrates the feasibility of searching for disks around young brown dwarfs 
at even lower masses with IRAC ($M<10$~$M_{\rm Jup}$).
In addition, photometry at longer wavelengths with the Multiband Imaging 
Photometer for Spitzer and spectroscopy with the Infrared Spectrograph on 
{\it Spitzer} of OTS~44 and brown dwarfs like it should provide detailed
constraints on the physical properties of disks around low-mass brown dwarfs, 
such as their mineralogy and geometry \citep[e.g.,][]{apa04,fur05}.

\acknowledgements
We are grateful to Robert Gutermuth for developing the software used
in reducing the IRAC images.
K. L. was supported by grant NAG5-11627 from the NASA Long-Term Space
Astrophysics program.
P. D. acknowledges grants from CONACyT and PAPIIT/DGAPA, M\'exico.
N. C. and L. H. acknowledge grant NAG5-13210.
This work is based on observations made with the {\it Spitzer Space Telescope},
which is operated by the Jet Propulsion Laboratory, California Institute 
of Technology under NASA contract 1407.
Support for this work was provided by NASA through contract 1256790 issued
by JPL/Caltech. Support for the IRAC instrument was provided by NASA through
contract 960541 issued by JPL. 
This publication makes use of data products from the Two Micron All
Sky Survey, which is a joint project of the University of Massachusetts
and the Infrared Processing and Analysis Center/California Institute
of Technology, funded by the National Aeronautics and Space
Administration and the National Science Foundation.

\clearpage

\begin{deluxetable}{lcccccccc}
\tabletypesize{\scriptsize}
\tablecaption{Photometry for OTS 44 and KPNO-Tau 4\label{tab:data}}
\tablehead{
\colhead{ID} &
\colhead{$I-J$} & 
\colhead{$J-H$} & 
\colhead{$H-K_s$} & 
\colhead{$K_s$} &
\colhead{$[3.6]$} &
\colhead{$[3.6]-[4.5]$} &
\colhead{$[4.5]-[5.8]$} &
\colhead{$[5.8]-[8.0]$}
} 
\startdata
OTS 44 & $\sim4.7$ & 1.01 & 0.79 & 14.61 & 13.67 & 0.41 & 0.53 & 0.70 \\
KPNO-Tau 4 & 3.73 & 0.97 & 0.74 & 13.28 & 12.51 & 0.10 & 0.12 & \nodata \\
\enddata
\end{deluxetable}

\clearpage

\end{document}